\documentclass[10pt, a4paper, oneside]{article}
\usepackage{cite}
\usepackage{graphicx}
\usepackage{subfigure}
\usepackage{epsfig}
\usepackage{epstopdf}
\usepackage{amssymb,amsmath}
\usepackage[a4paper=true,pagebackref=false]{hyperref}
\usepackage[scale=0.8]{geometry}
\hypersetup{colorlinks = true, linkcolor = cyan, anchorcolor = green, citecolor = red, filecolor = red, pagecolor = magenta, urlcolor = darkblue}

\title{Cosmic dynamics and qualitative study of Rastall model with spatial curvature}

\author{Ashutosh Singh${}^{1}$\footnote{ashuverse@gmail.com}
\\
${}^{1}$Department of Applied Mathematics, Jabalpur Engineering College,\\ Jabalpur, Madhya Pradesh, Pin 482011, India \\ \\ 
Gyan Prakash Singh${}^{2}$\footnote{gps.math2015@gmail.com} \\ ${}^{2}$Department of Mathematics, Visvesvaraya National Institute of Technology,\\ Nagpur, Maharashtra, Pin 440010, India\\ \\
Anirudh Pradhan${}^3$\footnote{pradhan.anirudh@gmail.com}\\
${}^3$Centre for Cosmology, Astrophysics and Space Science, GLA University,\\
	Mathura, Uttar Pradesh, Pin 281406, India\\
\\ }

\date{}
\begin{document}
\maketitle

\begin{abstract}
	We investigate the cosmic dynamics of Rastall gravity in non-flat Friedmann-Robertson-
	Walker (FRW) space-time with barotropic fluid. In this context, we are concerned about
	the class of model satisfying the affine equation of state. We derive the autonomous system
	for the Rastall model with barotropic fluid. We apply the derived system to investigate the
	critical points for expanding and bouncing cosmologies and their stable
	nature and cosmological properties. The expanding cosmology yields de Sitter universe
	at late times with decelerating past having matter and radiation dominated phases. Investigation of autonomous 
	system for bouncing cosmology yields oscillating solutions in positive spatial curvature region. 
	We also investigate the consequences of setting-up
	a model of non-singular universe by using energy conditions. The distinct features of
	the Rastall model compared to the standard model have been discussed in detail.
	
	\end{abstract}


\section{Introduction}
\label{introdution}
In Rastall's gravity theory, Rastall proposed generalizing the stress-energy conservation equation of Einstein’s theory 
but maintained the validity of the equivalence principle to make the divergence of energy-momentum tensor ($T_{ij}$) proportional to
the gradient of curvature scalar ($R$), that is, $T^{i}_{j;i} = \lambda R_{,j}$ where $\lambda$ denotes the Rastall
parameter \cite{1}. Relation $T^{i}_{j;i} = \lambda R_{,j}$ may be seen as result of quantum effects and these
effects may result in a `gravitational anomaly' for which the usual conservation of stress-energy tensor gets violated \cite{2,3}. 
One possible implication of conservation law in Rastall theory is to see the condition $T^{i}_{j;i}\neq 0$ as a consequence of particle creation
in cosmological context \cite{3,4,5,6,7}. The difficulty of Rastall's proposal is that the field equations are taken without any Lagrangian 
formulation. There have been various attempts for the Lagrangian formulation of Rastall theory in literature \cite{8,9,10,11}. Fabris
et al. \cite{11} have demonstrated that there is a way to cast the Rastall gravity theory with $f(R, T)$ theory \cite{12} is by the hydrodynamical 
representation of a scalar field. Rastall's theory provides a setting in which energy-momentum source and geometry can be non-minimally 
coupled to each other. Such mutual interaction of geometry and matter sources may lead to many interesting consequences. The energy density 
of the universe during inflation may be drawn from the negative energy of the gravitational source field \cite{13,14}. Various theoretical 
and observational features of Rastall theory have been studied in literature \cite{3,6,7,9,10,11,15,16,17,18,19,20,21,22,23}. In response to 
Visser’s claim that the Rastall gravity is completely equivalent to General Relativity \cite{24}, Darabi et al. \cite{20} have illustrated 
that Rastall gravity is a form of modified gravity different from General relativity.\\
In cosmological models, most of the governing equations are non-linear, and thus, dynamical system analysis may be used to describe the 
qualitative behavior of the model. In dynamical system analysis of governing equations of universe model, the critical points of the autonomous 
system govern different epochs of the universe. In this paper, we investigate the system of autonomous differential equations generated from 
the governing equations of Rastall theory. This method allows observation of the general behavior of the cosmological solutions associated with 
the present model. The dynamical system method has been applied to a wide class of models in different
gravity theories, for example see \cite{17,25,26,27,28,29,30,31,32,33,34,35,36,37,38,39,40,41,42,43,44} and references therein. It can be used to
obtain the class of solutions and their stability. These investigations may lead to interesting cosmological solutions, such as those 
with de Sitter phases or with radiation and matter-dominated phases. In Rastall's gravity theoretical framework with the quadratic equation of state 
signifying high energy state, the stability of bouncing solutions has been investigated by Silva et al. \cite{17}. By employing the dynamical system
technique, Khyllep and Dutta have found that at late times, the dynamics of the Rastall model in flat-FRW space-time may resemble the $\Lambda$ cold dark matter model \cite{33}. Cosmological implications of Rastall-$f(R)$ theory using dynamical system
technique in flat-FRW background predicts an accelerating universe for different values of Rastall parameter \cite{45}. The dynamical 
evolution of flat-FRW Rastall gravity model with linear equation of state yields oscillating solutions \cite{36}. In this paper, we
intend to analyze the qualitative behavior of the universe in Rastall theory by studying the critical points of the system. Since critical 
points of the dynamical system attract/repel trajectories for a wider parameter space around them, the system can be analyzed by studying 
dynamical evolution near the critical point. We concentrate on the dynamical evolution of the Rastall model with barotropic fluid by including 
spatial curvature. The observational evidence points to a (nearly) spatially flat universe, but this conclusion is based on a model with 
a restricted set of parameters \cite{46}. In cosmology, the possible features and properties of the present universe are studied. The non-minimally 
coupled nature of Rastall theory motivates to inspect the full phase of possibilities by studying the qualitative change in the dynamics of 
the universe by the inclusion of spatial curvature. We aim to investigate the implications of expanding cosmologies together with bouncing cosmologies 
by investigating the universe model in Rastall gravity.\\
The paper is organized as follows: in section 2, we write the cosmological equations in Rastall gravity for FRW space-time along with basic definitions. In section 3, we study the autonomous system of expanding cosmologies and highlight details on the
cosmological implications. In section 4, we study the bouncing cosmologies by using an autonomous system of governing equations with
cosmological implications. In section 5, we summarize the results.

\section{Cosmological model and equations}
In this section, we write the equations for Rastall gravity model. The FRW line element is
\begin{equation}
	ds^2=dt^2-a^2(t)\left(\frac{dr^2}{1-\kappa r^2} +r^2(d\theta^2+\sin^2\theta d\phi^2)\right)
	\label{eq1} 
\end{equation}
where $a(t)$ denotes the scale factor. The parameter $\kappa$ takes the values $0,+1,-1$ for
flat, closed and open spatial sections respectively. Field equations in Rastall gravity model may be given as\cite{1, 15}
\begin{equation}
	R_{ij}-\frac{1}{2}(1-2k\lambda)Rg_{ij}=kT_{ij}
	\label{eq2}
\end{equation}
where $g_{ij},R_{ij}$ denotes the metric tensor and Ricci tensor respectively. In the units $
G=c=1,k=8\pi$ and $\lambda=0$ will yield Einstein's equations of General theory of relativity, whenever $T^{ij}_{;j}=0$ \cite{1,15,23}. 
$k$ denotes the gravitational coupling constant and we consider it as a free parameter for modeling requirements with care to the
satisfaction of equations. In this model, we avoid $k\lambda = \frac{1}{4},\frac{1}{6}$, since some of the
quantities may be undefined at these values \cite{1,7,15, 23}.\\
The field equations in the Rastall gravity for perfect fluid source $T_{ij}=\text{diag}(\rho,-p,-p,-p)$ may be given by \cite{19,23}
\begin{equation}
	3(1-4k\lambda)H^2-6k\lambda \dot{H}+3(1-2k\lambda)\frac{\kappa}{a^2}=k\rho
	\label{eq3}
\end{equation}
\begin{equation}
	3(1-4k\lambda)H^2+2(1-3k\lambda) \dot{H}+(1-6k\lambda)\frac{\kappa}{a^2}=-kp
	\label{eq4}
\end{equation}
where $\rho, p$ denotes the energy density and pressure. The Hubble parameter is given by $H=\frac{\dot{a}}{a}$. Overhead dot denotes 
derivative with respect to cosmic time $t$. The Bianchi identity ($G^{;j}_{ij}=0$) leads to the continuity equation as \cite{15}
\begin{equation}
	(3k\lambda-1)\dot{\rho}+3k\lambda \dot{p}+(4k\lambda-1)3H(\rho+p)=0
	\label{eq5}
\end{equation}
In the present model, we take the barotropic equation of state $p=f(\rho)$ for the perfect fluid in our present setting and keep our
discussion general for the moment. The nature of critical points in dynamical system analysis can be examined by studying the linear
deviations around the critical points. The linear deviations follow the equation $X_i^{'}= MX_j$, where $M$ denotes the Jacobian matrix
of the linearized system. The eigenvalues $\lambda_i$ of $M$ calculated for each critical point highlight
the stable nature of critical points. If the real part of eigenvalues of matrix $M$
are non-zero, then the point is called hyperbolic \cite{36,47}. A hyperbolic point is a sink
(or attractor) and source (or repeller) if real parts of all eigenvalues are negative and positive, respectively. In the case of attractors,
the perturbed system around them
will return to the equilibrium state. On the other hand, the system will not return to the equilibrium state in the case of repellers.
If eigenvalues are negative as well as positive, then the hyperbolic point is said to be a saddle point \cite{37,38}.\\
Phase space analysis and diagrams are very convenient tools for visualizing oscillatory systems. The trajectories
can reach neutrally stable states known as centers in these systems. For centers, the eigenvalues of the Jacobian matrix at critical points 
are purely imaginary and conjugated \cite{28,35,39}.\\ 
In cosmological modeling, energy conditions may categorize certain physically reasonable ideas in a precise manner. The point-wise energy
conditions are the coordinate invariant restrictions on the energy-momentum tensor of the model. These energy conditions are termed Null energy
condition (NEC), Weak energy condition (WEC), Dominant energy condition (DEC), Strong energy condition (SEC) and may be written as \cite{23}
\begin{enumerate}
	\item NEC $\Leftrightarrow \rho+p\geq 0$
	\item WEC $\Leftrightarrow \rho\geq 0, \rho+p\geq 0$
	\item DEC $\Leftrightarrow \rho\geq 0,\rho\pm p\geq 0$
	\item SEC $\Leftrightarrow \rho+p\geq 0,\rho+3p\geq 0$
\end{enumerate}
These conditions exhibit various combinations of energy density and pressure of the model to be non-negative. These conditions can 
be compatible with cosmic acceleration in models provided that a component yielding repulsive gravity exists and acceleration stays within 
certain bounds.
\section{Expanding cosmology}
\begin{figure}[h!]
	\centering
	\subfigure[]{\includegraphics[width=0.325\textwidth]{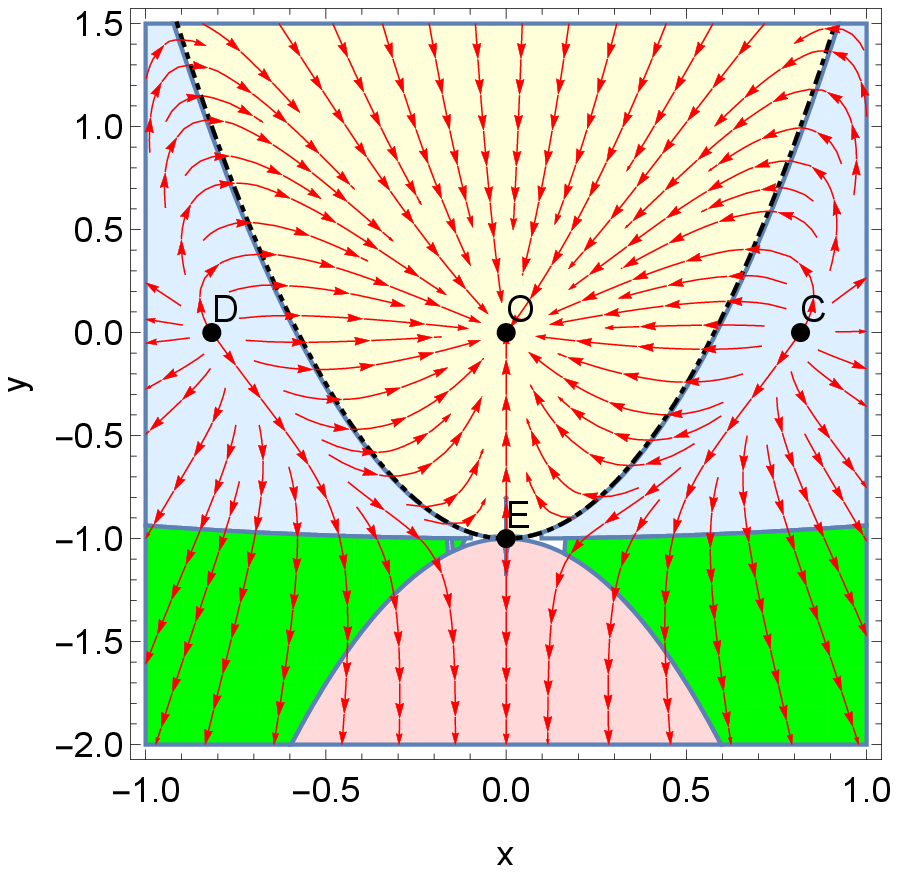}} 
	\subfigure[]{\includegraphics[width=0.325\textwidth]{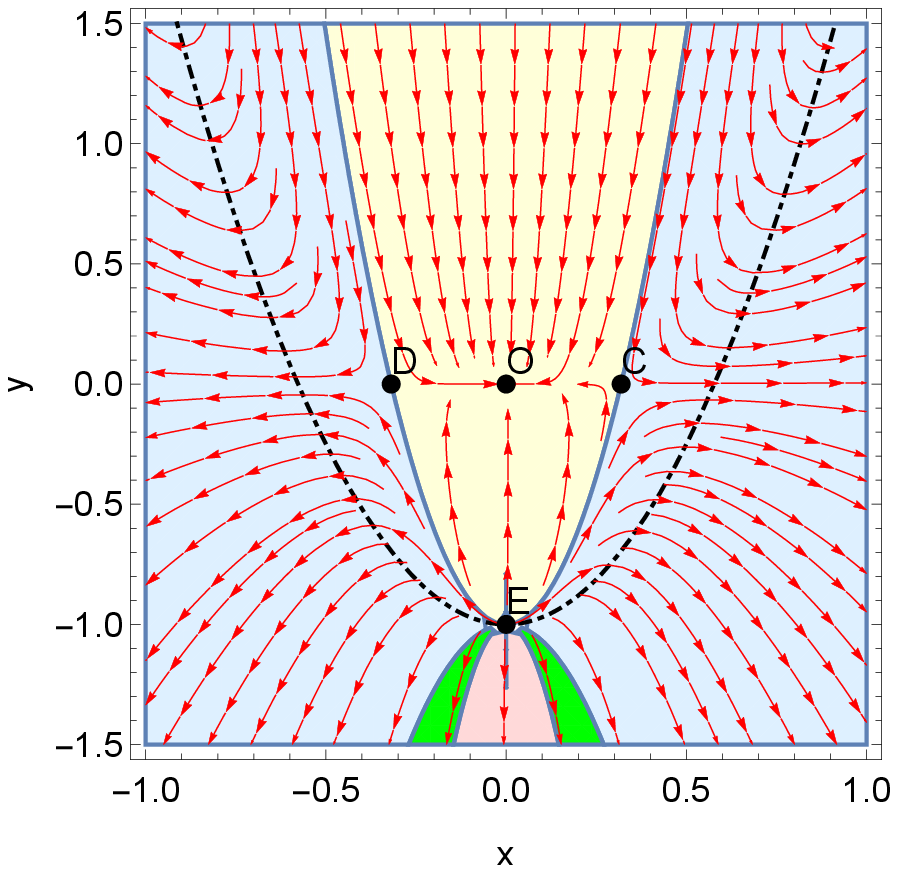}} 
	\subfigure[]{\includegraphics[width=0.325\textwidth]{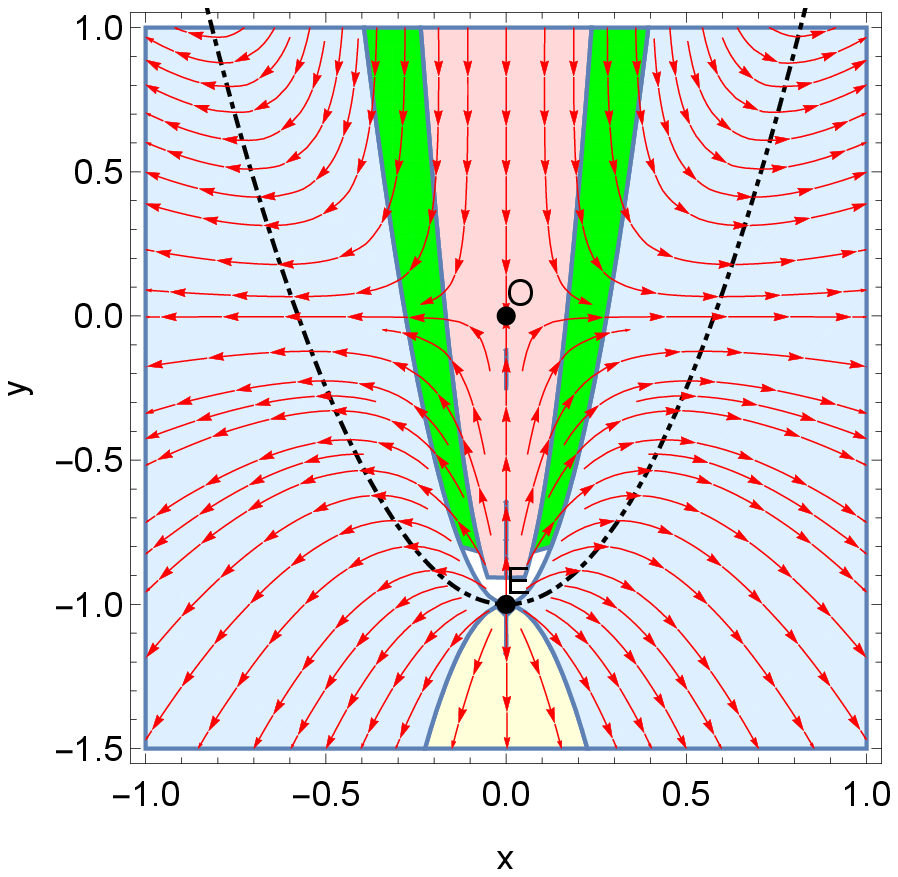}}
	\caption{Different views of phase space $(x,y)$ for (a) $A<-\frac{2}{3}$ (b) $-\frac{2}{3}<A<0$ (c) $A>0$ respectively}
	\label{fig1}
\end{figure}

\begin{figure}[h!]
	\centering
	\subfigure[]{\includegraphics[width=0.4\textwidth]{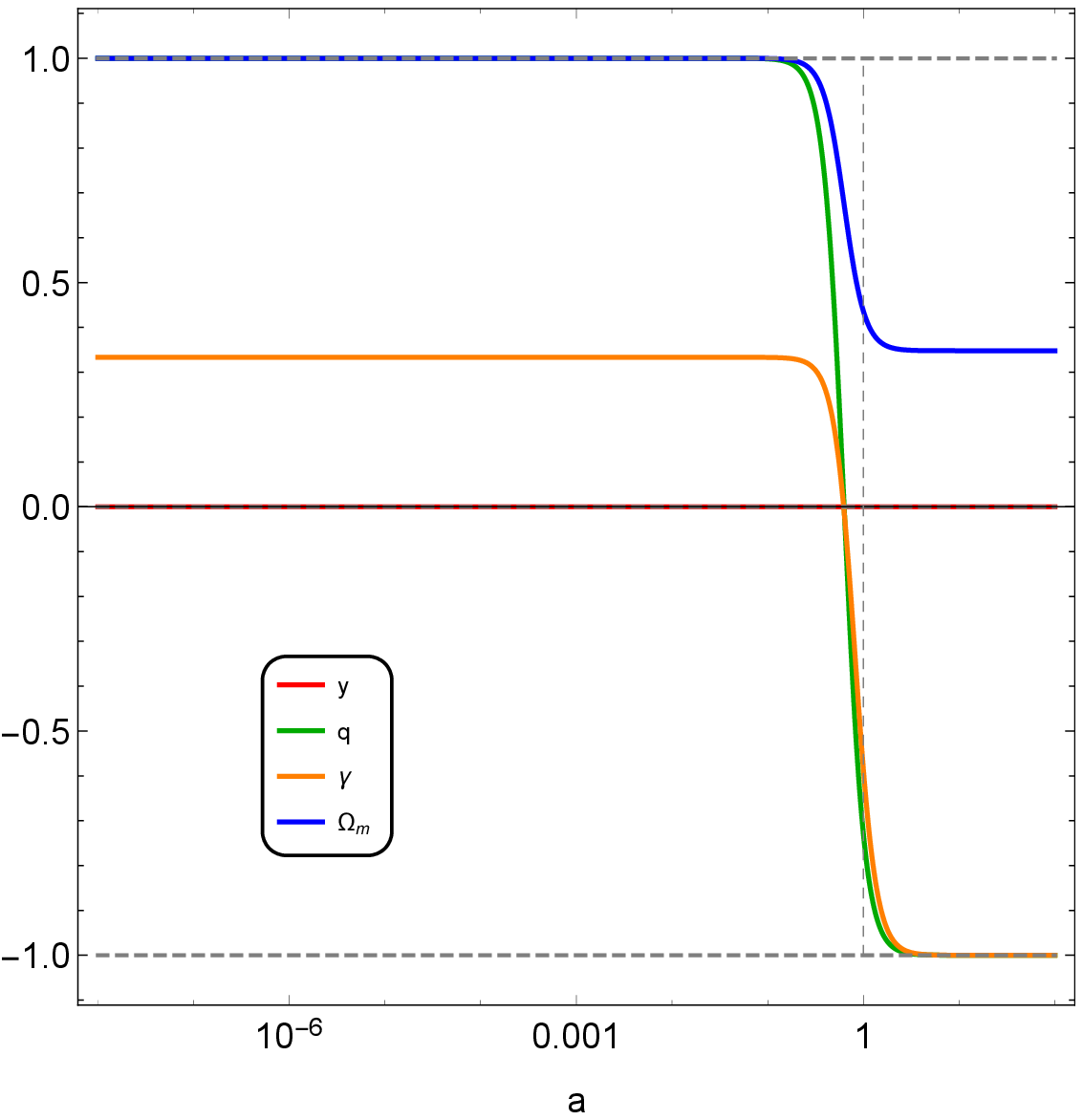}} \hfill
	\subfigure[]{\includegraphics[width=0.4\textwidth]{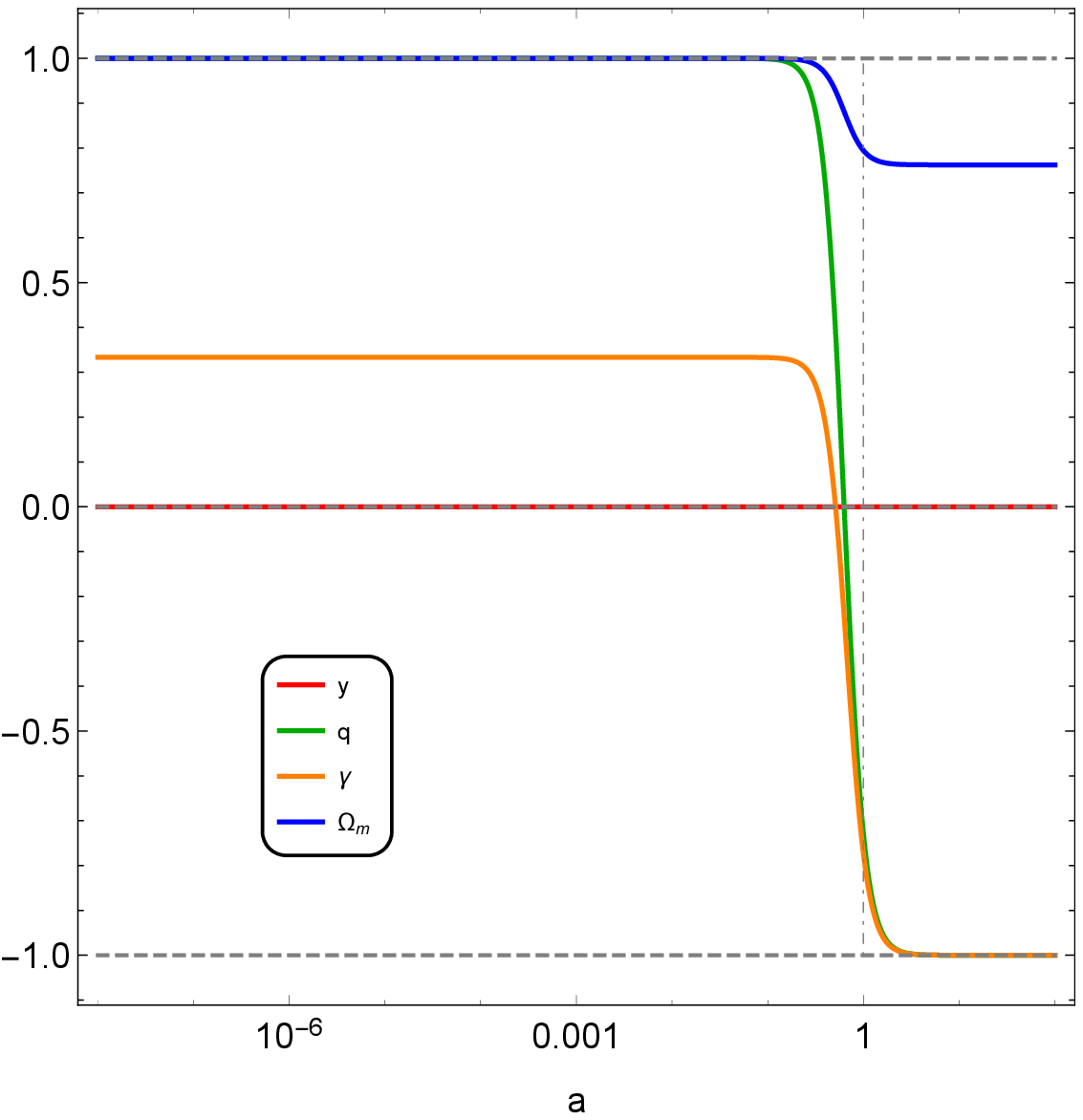}} 
	\subfigure[]{\includegraphics[width=0.4\textwidth]{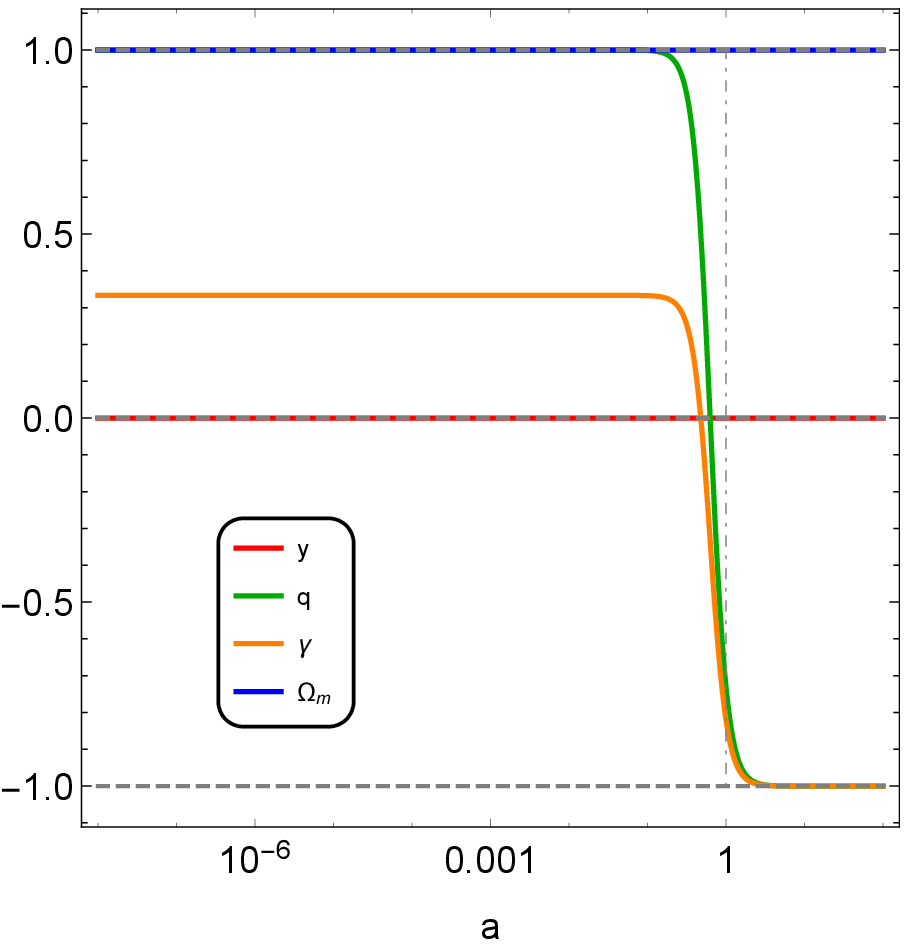}}\hfill
	\subfigure[]{\includegraphics[width=0.4\textwidth]{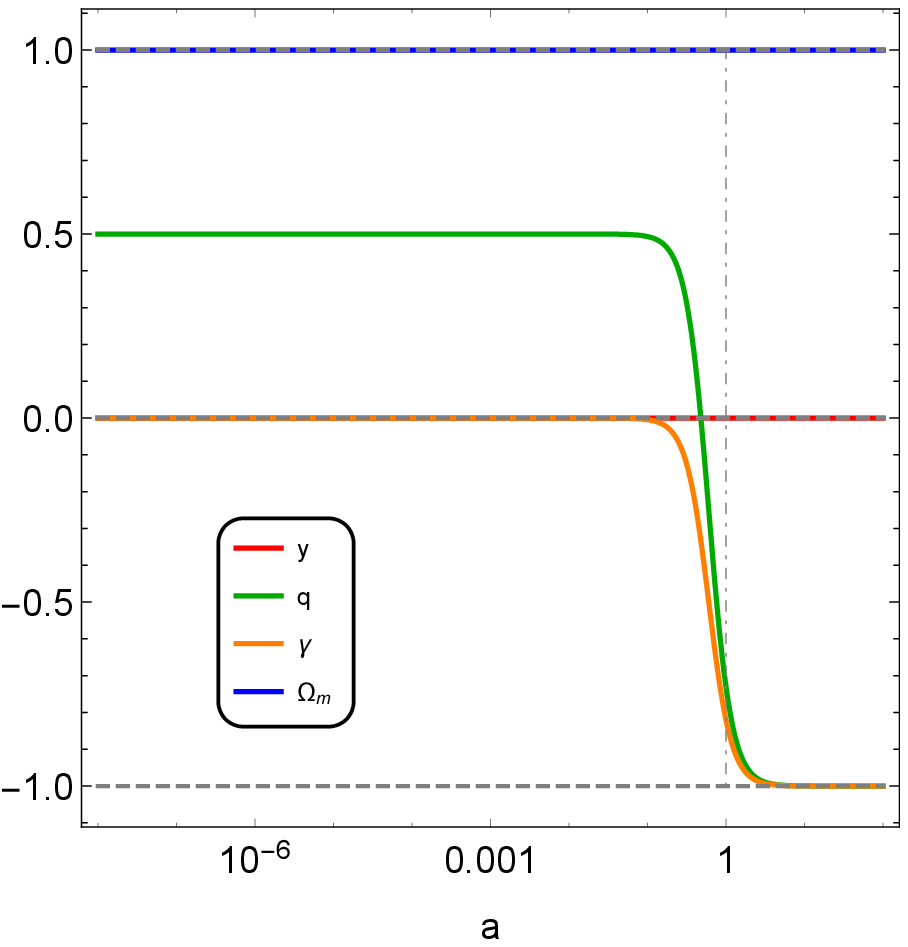}}
	
	\caption{$q,\gamma,\Omega_m$ and normalized curvature with $a$ for model parameter values 
		(a) $\alpha=\frac{1}{3},k=1,\lambda=0.163$, (b) $\alpha=\frac{1}{3},k=1,\lambda=0.06$, (c) $\alpha=\frac{1}{3},k=1,\lambda=0.00005$, 
		(d) $\alpha=0,k=1,\lambda=0.00005$ \label{fig2}}
\end{figure}

\begin{figure}[h!]
	\centering
	\subfigure[]{\includegraphics[width=0.325\textwidth]{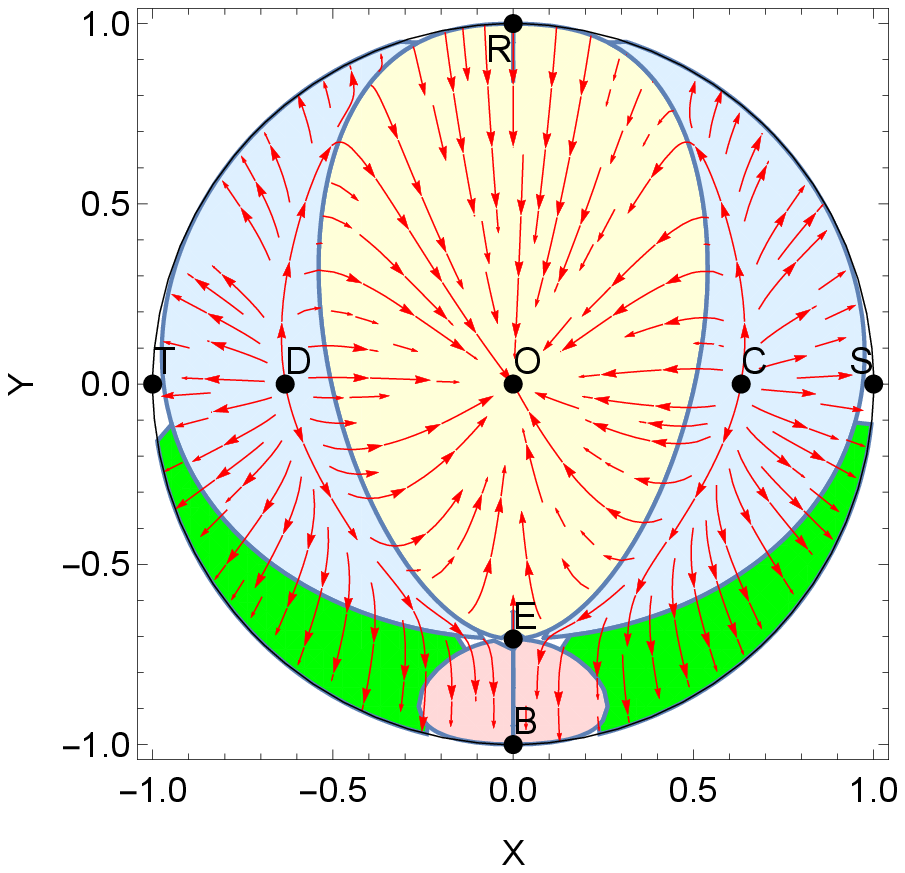}} 
	\subfigure[]{\includegraphics[width=0.325\textwidth]{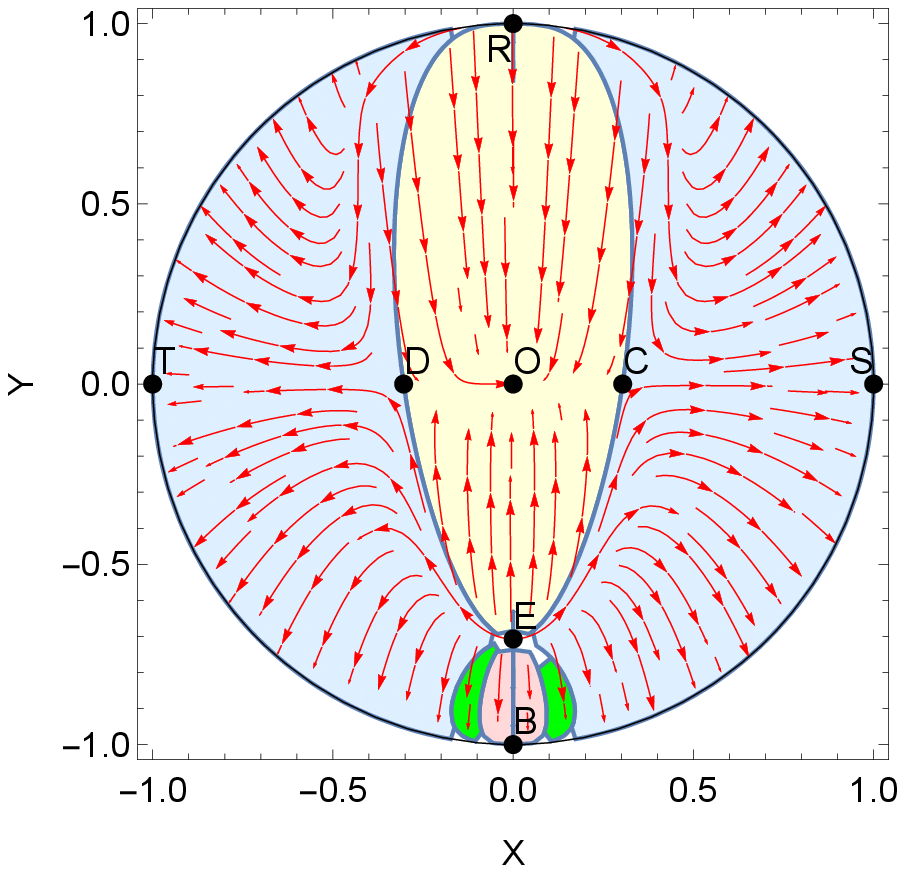}} 
	\subfigure[]{\includegraphics[width=0.325\textwidth]{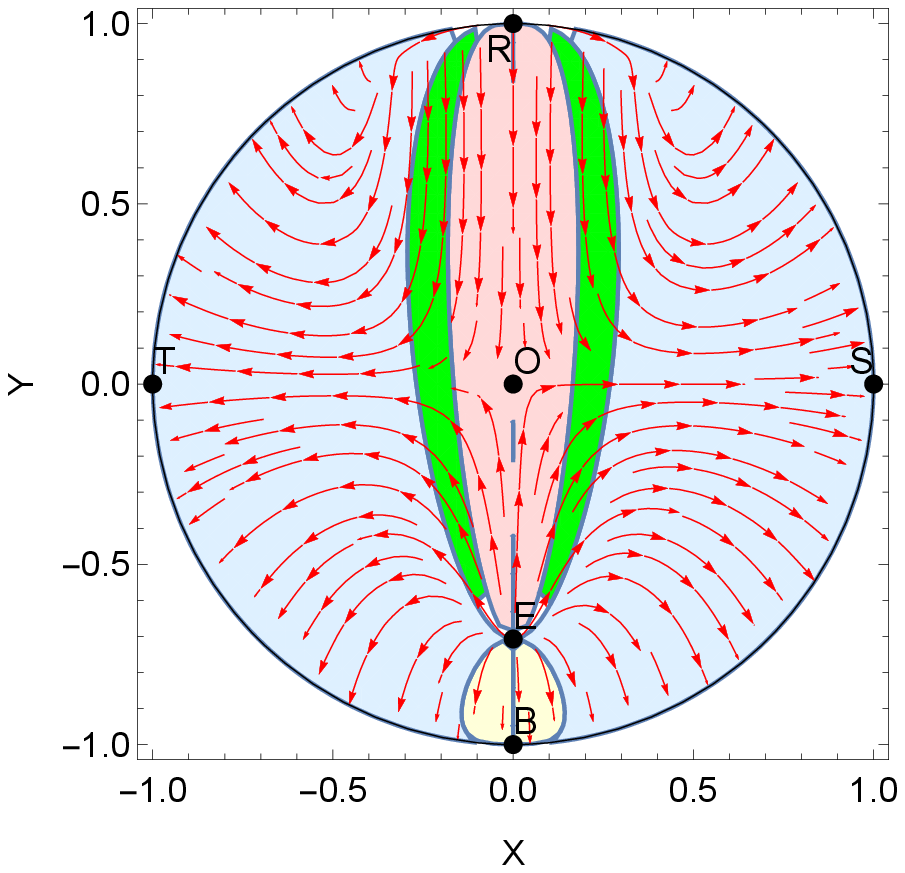}}
	\caption{Different views of phase space $(X,Y)$ for (a) $A<-\frac{2}{3}$ (b) $-\frac{2}{3}<A<0$ (c) $A>0$ respectively}
	\label{fig3}
\end{figure}

In this section, we study the governing equations of model via dynamical system method. In analogy with scalar field models, we introduce the 
following notations \cite{25}
\begin{equation}
	\rho_k\equiv \frac{\rho+p}{2},\quad p_v\equiv \frac{\rho-p}{2}
	\label{eq6}
\end{equation}
The autonomous system may be formulated by defining the variables
\begin{equation}
	x^2=\frac{k\rho_k}{3H^2}, \quad y=\frac{\kappa}{a^2H^2}, \quad z^2=\frac{k\rho_v}{3H^2}
	\label{eq7}
\end{equation}
To study the qualitative aspects, we take the derivative of these variables concerning the number of e-folding $N=\ln a$, thus $dN=Hdt$.
We may anticipate that the above system will display accelerated exponential expansion from this formulation. The connection between 
curvature and dynamics becomes more explicit with the above parametrization of governing equations. In terms of the above variables, 
we may define the curvature density parameter as $\Omega_k=-y$. Zero spatial curvature models correspond to $y=0$. The variable $y$ 
cannot change the sign in our model. The model with positive (negative) curvature will have $y>0$, ($y<0$)
and no trajectories in the phase plane can cross $y>0$ to $y<0$ or vice-versa.\\ 
For the variables (\ref{eq7}), the autonomous system takes the form
\begin{equation}
	x'=x\left(3x^2-y+\frac{3A}{2} \right)
	\label{eq8} 
\end{equation}
\begin{equation}
	y'=2y\left(3x^2-y-1\right)
	\label{eq9} 
\end{equation}
\begin{equation}
	z'=\frac{1-f'(\rho)}{1+f'(\rho)}\frac{3A}{2}\frac{x^2}{z}+z\left(3x^2-y\right)
	\label{eq10} 
\end{equation}
where we define $f'(\rho)\equiv\frac{df}{d\rho}$ and $A\equiv \frac{(1+f'(\rho))(1-4k\lambda)}{3k\lambda(1+f'(\rho))-1}$. 
From Eq. (\ref{eq3}) and (\ref{eq4}), we may write $\frac{\dot{H}}{H^2}=-3x^2+y$ and $\frac{k\rho}{3H^2}=(1-4k\lambda)+6k\lambda x^2+(1-4k\lambda) y$. 
By defining $\Omega_m=\frac{k\rho}{3H^2}\geq 0$, we have $\frac{6k\lambda}{(4k\lambda-1)}x^2+y\leq 1$. The above system of 
Eq. (\ref{eq8},\ref{eq9},\ref{eq10}) may be used to study the qualitative behavior of models with barotropic fluid.\\
On the physical grounds in cosmological modeling, a functional dependence of energy density on the pressure of the model may be postulated. 
We consider a perfect fluid with affine equation of state having form \cite{26}
\begin{equation}
	p=\alpha\rho-\rho_0
	\label{eq11}
\end{equation}
here $\alpha, \rho_0$ are free parameters. This affine equation of state (EoS) provides the description of both hydrodynamically 
unstable $(\alpha < 0)$ and stable $(\alpha > 0)$ fluids. This affine EoS may also be seen as generalization of $p=\alpha\rho$, 
($\alpha=$ constant) which is hydro-dynamically unstable for $\alpha < 0$. This simple generalization allow to include
the dark (and also phantom) energy with $\alpha\geq 0$. With the use of linear EoS in the model, we may have negative pressure in 
the model even with $\alpha>0$. In this model, squared sound speed of linear perturbations is given by 
${c_s}^2=\frac{\partial p}{\partial \rho}=\alpha$. The model is classically stable for $0 \leq \alpha\leq 1$. In this paper, we are 
using classical stability criterion from Eq. (\ref{eq11}). The EoS parameter $(\gamma)$ is given by $\gamma=\frac{p}{\rho}$. 
NEC is said to be valid for $\rho+p\geq 0\Rightarrow \rho\geq \frac{\rho_0}{1+\alpha}$ \cite{36}. The fluid violates NEC in the phantom region. 
This condition simply means that the energy density during the evolution of the universe in the model will be increasing and decreasing for expanding
and contracting scenario respectively \cite{25}. From Eq. (\ref{eq5}) and (\ref{eq11}), we have
\begin{equation}
	\rho=\frac{a_0}{1+\alpha}a^{3A}+\frac{\rho_0}{1+\alpha},\quad \alpha\neq -1 \nonumber
\end{equation}
\begin{equation}
	\rho=\log\left(\frac{a_0}{a^{3\rho_0(4k\lambda-1)}} \right),\quad \alpha= -1 \nonumber
\end{equation}
From Eq. (\ref{eq3}), (\ref{eq4}) and (\ref{eq7}), we may write $\frac{6k\lambda-1}{4k\lambda-1}x^2-y-\frac{1}{4k\lambda-1}z^2=1$, so, the phase 
space defined through Eq. (\ref{eq8},\ref{eq9},\ref{eq10}) reduces into two-dimensional system and by using Eq. (\ref{eq11}), we have
\begin{equation}
	x'=x\left(3x^2-y+\frac{3A}{2} \right)
	\label{eq12} 
\end{equation}
\begin{equation}
	y'=2y\left(3x^2-y-1\right)
	\label{eq13} 
\end{equation}
where $f'(\rho)=\alpha$. We study the structure of dynamical equations by using phase space analysis of variables $x$ and $y$. Critical points
may be calculated by setting $x'=0$ and $y'=0$. This autonomous system may exhibit interesting scenarios based on critical points. 
Recall that in this section, we exclude $k\lambda=\frac{1}{4},\frac{1}{6}$ and $k\lambda=\frac{1}{3(1+\alpha)}$. 
We have $A\equiv \frac{(1+\alpha)(1-4k\lambda)}{3k\lambda(1+\alpha)-1}$ and it may be greater than or less than $0$
depending on the value of $\alpha$ and $k\lambda$. For $\alpha=-1$ or $k\lambda=\frac{1}{4}$, we may have $A=0$.
Here, $k\lambda\neq \frac{1}{4}$ so $\alpha=-1$ for $A=0$. For $A<0$, we may have either of the following
conditions on $\alpha$ and $k\lambda$:
\begin{enumerate}
	\item $\alpha<-1$ and $k\lambda>\frac{1}{4}$ or $k\lambda<\frac{1}{3(1+\alpha)}$
	\item $-1<\alpha\leq \frac{1}{3}$ and $k\lambda<\frac{1}{4}$ or $k\lambda>\frac{1}{3(1+\alpha)}$
	\item $\alpha>\frac{1}{3}$ and $k\lambda>\frac{1}{4}$ or $k\lambda<\frac{1}{3(1+\alpha)}$
\end{enumerate}
For $A>0$, we will have complementary conditions of above stated conditions. In cosmology, the deceleration parameter highlights the 
rate at which universe expansion is slowing down and is defined in terms of Hubble parameter as $q=-1-\frac{\dot{H}}{H^2}$. For $q<-1,q=-1$ 
and $-1<q<0$, universe will undergo super-exponential, de Sitter and accelerated expansion respectively. Eternal acceleration is achieved when 
$q<0$ for all time and the universe decelerates for $q>0$\cite{48,49,50,51}. For variables $x,y$ (see Eq. (\ref{eq7})), deceleration parameter 
$(q)$ and effective EoS parameter $(\gamma)$ are given by
\begin{equation}
	q=3x^2-y-1,\quad \gamma=-1+\frac{2x^2}{6k\lambda x^2+(1-4k\lambda)(1+y)}
	\label{eq14}
\end{equation}
For the autonomous system (\ref{eq12},\ref{eq13}), there will be four critical points in finite
space given by $O(x=0,y=0), C(x=-\Xi,y=0),D(x=\Xi,y=0),E(x=0,y=-1)$ where we denote 
$\Xi=\frac{\sqrt{(1+\alpha)(4k\lambda-1)}}{\sqrt{6k\lambda(1+\alpha)-2}}$. Various detail about these critical points and geometrical 
behavior in the $x-y$ plane has been summarized in Table (\ref{table1}) and Fig. (\ref{fig1}). In figures, the arrows represent direction of velocity
field and the trajectories reveal their stability nature. Also, in phase space diagrams $\gamma<-1,-1<\gamma<0,0<\gamma<1$ and $\gamma>1$ 
region have been highlighted by light-red, light-yellow, light-blue, green shades respectively. For $(\alpha,k,\lambda,A)$ equals to 
$\left( \frac{1}{3},1,0.163,-1.3238\right)$, $\left( 0,1,0.235,-0.20339\right)$ and $\left( 0.15,1,0.261,0.50828\right)$, we
plot left, middle and right panel of Fig. (\ref{fig1}) respectively.\\
The Rastall coupling constant $k$ and Rastall constant parameter $\lambda$ have been constrained at different scales of the universe 
	in Rastall gravity literature. Li et al. \cite{51a} have given the value of $k\lambda=0.163\pm 0.001$ ($68\%$ CL) by using methods based 
	on galaxy-galaxy strong gravitational lensing system. Akarsu et al. \cite{51b} have found that $\lambda$ (with $k=1$) can be in range 
	$-0.0007<\lambda<0.0001$ at $68\%$ CL from CMB+BAO data. We have used these constrained values for illustration of the phase portraits 
	and cosmological parameters in Fig. (\ref{fig1},\ref{fig2},\ref{fig3}). These selected values of parameters give us the desired values 
	for the combined parameter $A$ (namely $A<-\frac{2}{3}$, $-\frac{2}{3}<A<0$, $A>0$) to display the topologically in-equivalent phase portraits. 
	These inequalities involving $A$ give a large parameter space for $k\lambda$ and $\alpha$. However, the value of $\alpha$ is 
	constrained by using the classical stability criterion $0\leq \alpha\leq 1$. In terms of dynamical system constraints, these classes of 
	values of $A$ give different behavior of critical points (in terms of stability). However, by assisting with the observational bounds 
	of $\lambda$ from Li et al.\cite{51a} and Akarsu et al.\cite{51b}, we may see that the value of $\lambda$ is affecting the cosmic dynamics 
	of parameters like $\gamma$ and $q$ (see Fig. (\ref{fig2})). From the qualitative analysis, we may observe that to maintain the positive matter density value, we should restrict 
	$\lambda>0$ for $k>0$ cases. \\
	The barotropic equation of state (\ref{eq11}) reduces to $p=\alpha\rho$ for $\rho_0=0$. In the present formulation of dynamical variables, 
	the system uses $f'(\rho)$ term and thus $p=\alpha\rho-\rho_0$ and $p=\alpha\rho$ contributes only with $\alpha$ in the dynamical system 
	(see Eq. (\ref{eq12}) and (\ref{eq13})). However, one may note that $\rho_0$ will affect the evolution of energy density and pressure of 
	the model. In order to display the evolution of dark matter dominated universe along-with radiation dominated universe, we have taken 
	$\alpha=0$ and $\alpha=\frac{1}{3}$ respectively. The first and second panel of Fig. (\ref{fig1}) and (\ref{fig3}) have been using 
	$\alpha=\frac{1}{3}$ and $\alpha=0$ respectively. A large combination of values of $k\lambda$ and $\alpha$ may be taken but, these will 
	yield three kinds of topologically different phase portraits in variables $(x,y)$ or $(X,Y)$. However, the region of $\gamma$ in the phase 
	portraits may change by changing the values of $k\lambda$ and $\alpha$.\\
	The behavior of EoS parameter along with other cosmic parameters for $\alpha=\frac{1}{3},k=1$ and $\lambda=0.163$\cite{51a}, 
	$\lambda=0.06$,\cite{51b} $\lambda=0.00005$ \ \cite{51b} have been displayed in first, second and third panel of Fig. \ref{fig2}. 
	The fourth panel uses the value $\alpha=0,k=1,\lambda=0.00005$. This shows that the model may reveal the late-time accelerating universe. 
	The model may yield EoS parameter $\gamma$ value compatible with observations $\gamma=-0.957\pm0.080$ at the present time\cite{46} for 
	the model parameter values $\alpha=\frac{1}{3},k=1$ and $\lambda=0.00005$, which are consistent with constraints of Akarsu et al.\cite{51b} also. 
	It may be concluded that the Rastall parameter value of order $10^{-4}$ may also yield the late-time accelerating universe with the matter 
	density taking value close to unity at present times. \\
\begin{table}[h!]
	\begin{center}
		{\begin{tabular}{ccccc}
		\hline\noalign{\smallskip}
		Point & $\lambda_1$ & $\lambda_2$ &
			$q$ & $\Omega_m$ \\
			\noalign{\smallskip}\hline\noalign{\smallskip}
			O & $-2$ & $\frac{3A}{2}$ & $-1$ & $1-4k\lambda$\\
			C & $-2-3A$ & $-3A$ & $-1-\frac{3A}{2}$ & $1-4k\lambda-3k\lambda A$\\
			D & $-2-3A$ & $-3A$ & $-1-\frac{3A}{2}$ & $1-4k\lambda-3k\lambda A$\\
			E & $2$ & $\frac{1}{2}(2+3A)$ & $0$ & $0$\\
			\noalign{\smallskip}\hline
		\end{tabular}
	\caption{Eigenvalues and quantities $q,\Omega_m$ for critical points of expanding cosmologies}
\label{table1}}      
\end{center}
\end{table}
The EoS parameter satisfying $\gamma=-1$ is an interesting case from the cosmological point of view due 
to the possibility of describing the dark energy or inflationary phenomena. Since $\frac{\dot{H}}{H^2}=-3x^2+y$, so, it may integrated at 
point $(x_0,y_0)$  of the phase space to give $a\propto (t-t_0)^{\frac{1}{3{x_0}^2-y_0}}$. The value of $a(t)$ only holds if the system 
is at one of the critical/fixed points. For other cases, one needs to numerically integrate the equation $\frac{\dot{H}}{H^2}=-3x^2+y$, 
since its right-hand side quantity may not be a constant. The behavior of the critical points in the phase space may be summarized as follows:\\
\textit{Point O}: This point will exist for all values of $\alpha$ and $k\lambda$. The point is either an attractor
(for $A<0$) or saddle (for $A>0$). For the $A<0$ case, this point stands for dark energy dominated solution having effective EoS $\gamma=-1$ 
which exhibits an accelerating de Sitter expansion scenario and thus serves as a late-time attractor in the model.\\
\textit{Point C and D}: These points will exist only for $A<0$. But these points will be either saddle for $A\in \left(-\frac{2}{3},0 \right)$ 
or repeller for $A<-\frac{2}{3}$. These points exhibit accelerated as well as decelerated expansion with effective EoS depending on parameter 
of classical stability $\alpha$. For $A=0$, the points exhibit de Sitter accelerated expansion. For super-accelerated expansion $(q<-1)$, 
we need $A>0$. Therefore, these point will never exhibit super-accelerated expansion. With $A<0$, we get $q>-1$. In
particular, $q<0$ for $A\in \left(-\frac{2}{3},0 \right)$, $q=0$ at $A=-\frac{2}{3}$ and $q>0$ for $A<-\frac{2}{3}$. Thus, we may conclude 
that these points will undergo accelerated expansion with saddle nature and decelerated expansion with repeller nature. The black curve 
in Fig. (\ref{fig1}) corresponds to $q=0$ and region inside it represents an accelerating region of the phase space in the model.\\
\textit{Point }E: This point will exist for all values of $\alpha$ and $k\lambda$. The point is repeller for $A>-\frac{2}{3}$ and saddle 
for $A<-\frac{2}{3}$. This point lie on $q=0$ curve and in small neighborhood
of this point $p\approx -\rho$. At this point, $a\propto (t-t_0)$. This point represents Milne solution with $q=0$.
\subsection{Analysis of critical points at infinity}
To understand the global dynamics of a dynamical system, we may utilize Poincare compactification method. The phase space in our model 
is having variables $x,y$ and we denote $P=x\left(3x^2-y+\frac{3A}{2} \right)$ and $Q=y(6x^2-2y-2)$. By following ing the method 
stated in the Appendix (\ref{append}), we have $P^{*}(X,Y,Z)=X\left(3X^2-YZ+\frac{3A}{2}Z^2 \right)$ and $Q=Y(6X^2-2YZ-2Z^2)$. 
In this case $m=3$. Under the limit $Z\rightarrow 0$, $XQ_m-YP_m=3X^3Y$ along with $X^2+Y^2=1$. So, the equilibrium (fixed) 
points are $(\pm 1,0)$ and $(0,\pm 1)$. The phase space in variables $(X,Y)$ has been given in Fig. (\ref{fig3}). In the Poincare 
phase space, we have four points of finite space along with the points at infinity. The point $O$ is the late-time attractor of model. 
We may use fan-out maps to study the flow near equilibrium, along the circle of infinity \cite{47} for $X>0$ and $Y>0$ respectively. 
At points $(\pm 1,0)$ the deceleration parameter is $-1$ and $\gamma=-1+\frac{1}{3k\lambda}$. In order to study an equilibrium at 
circle of infinity $Z=0$
satisfying $X>0$, we may map a point $(X,Y,Z)$ to $U=\frac{Y}{X},V=\frac{Z}{X}$. The differential
equation governing the motion are
\begin{equation}
	U'=3U,\quad V'=-3V
	\label{eq15}
\end{equation}
The equilibrium is at $(U,V)=(0,0)$ and eigenvalues at this point are $\lambda_1=3,\lambda_2=-3$. The point is saddle in nature. 
The equilibrium point at $Z=0$ satisfying $X<0$ will also have a saddle nature. The points $(\pm 1,0)$ are saddle in nature and for 
suitable values of $k\lambda<\frac{1}{3}$, we may have matter and radiation dominated universe. And, at point $(0,\pm 1)$, the deceleration 
parameter is undetermined and becomes
indeterminate. Consider an equilibrium at $Z=0$ satisfying $Y>0$. A point $(X,Y,Z)$ is mapped to $U=\frac{X}{Y}$, $V=\frac{Z}{Y}$. 
The differential equation governing the motion is
\begin{equation}
	U'=-3U^3,\quad V'=-6U^2V
	\label{eq16}
\end{equation}
The equilibrium is at $(U,V)=(0,0)$ and eigenvalues at this point are $\lambda_1=0,\lambda_2=0$. The equilibrium point at $Z=0$ 
satisfying $Y<0$ will also have similar nature. From Fig. (\ref{fig3}), we may observe that for selected values of model parameters, 
these points are saddle in nature. Due to the indeterminate nature of $q$ and $\gamma$, these points are not of physical interest.

\section{Bouncing cosmology}
The bouncing universe undergoes a collapse, attains a minimum, and then subsequently expands \cite{52,53}. For a bounce to happen, 
we need $\dot{a}<0$ to $\dot{a}>0$ with $\dot{a}_b=0$ and $\ddot{a}>0$ in a small neighborhood of bounce point $t=t_b$. The first condition
of bounce may be expressed in terms of the Hubble parameter as $H(t=t_b)=0$, but this means that the dynamical variables defined in 
Eq. (\ref{eq7}) will diverge. These complications arise because the dynamical variables in Eq. (\ref{eq7}) were chosen for the study of 
expanding cosmological scenarios (that is, for $H>0$ scenarios only). To obtain the autonomous system for the non-singular cosmological 
scenario, we define the dynamical variables as:
\begin{equation}
	\bar{x}^2=\frac{3H^2}{k\rho_k}, \quad \bar{y}=\frac{\kappa}{k a^2\rho_k}, \quad \bar{z}^2=\frac{\rho_v}{\rho_k}
	\label{eq17}
\end{equation}
where $\rho_k\equiv \frac{\rho+p}{2},\quad p_v\equiv \frac{\rho-p}{2}$. To study the qualitative aspects, we define the new independent 
variable defined in analogy to $N$ as $\bar{N}=\int\sqrt{\frac{k\mid\rho_k\mid}{3}}dt$, that is $d\bar{N}=\sqrt{\frac{k\mid\rho_k\mid}{3}}dt$. 
The above variables are well behaved only for $\rho_k\neq 0$ and in this paper, we take $\rho_k\neq 0$. For the variables (\ref{eq17}), 
the autonomous system takes the form
\begin{equation}
	\bar{x}'=-\frac{3A}{2}{\bar{x}}^2+3\bar{y}-3 
	\label{eq18} 
\end{equation}
\begin{equation}
	\bar{y}'=-(2+3A)\bar{x}\bar{y}
	\label{eq19} 
\end{equation}
\begin{equation}
	\bar{z}'=\frac{1-f'(\rho)}{1+f'(\rho)}\frac{3A}{2}\frac{\bar{x}}{\bar{z}}-\frac{3A}{2}\bar{x}\bar{z}
	\label{eq20} 
\end{equation}
where we define $f'(\rho)\equiv\frac{df}{d\rho}$ and $A\equiv \frac{(1+f'(\rho))(1-4k\lambda)}{3k\lambda(1+f'(\rho))-1}$. From Eq. (\ref{eq3}) and (\ref{eq4}), we may write $\frac{\dot{H}}{k\rho_k}=\bar{y}-1$. The variables $\bar{x},\bar{y},\bar{z}$ are not 
independent and related by $\frac{4k\lambda-1}{6k\lambda-1}\bar{x}^2+3\frac{4k\lambda-1}{6k\lambda-1}\bar{y}+\frac{1}{6k\lambda-1}\bar{z}^2=1$. 
Above Eq. (\ref{eq18},\ref{eq19},\ref{eq20}) may be used to study the autonomous system of Rastall model with barotropic fluid $p=f(\rho)$. In the
following discussion, we take $f(\rho)=\alpha\rho-\rho_0$ so $f'(\rho)=\alpha$. For the variables (\ref{eq17}),
the resulting two-dimensional autonomous system takes the form:
\begin{equation}
	\bar{x}'=-\frac{3A}{2}{\bar{x}}^2+3\bar{y}-3 
	\label{eq21} 
\end{equation}
\begin{equation}
	\bar{y}'=-(2+3A)\bar{x}\bar{y}
	\label{eq22} 
\end{equation}
In terms of variables $\bar{x},\bar{y}$, the parameters $q$ and $\gamma$ are given by
\begin{equation}
	q=-1-\frac{3(\bar{y}-1)}{{\bar{x}}^2},\quad \gamma=-1+\frac{2}{6k\lambda+(1-4k\lambda)({\bar{x}}^2+3\bar{y})}
	\label{eq23}
\end{equation}
\begin{table}[h!]
	\begin{center}
		{\begin{tabular}{ccccc}
				\hline\noalign{\smallskip}
			Point & $\lambda_1$ & $\lambda_2$ &
			$\gamma$ & $q$ \\
			\noalign{\smallskip}\hline\noalign{\smallskip}
			H & $3\sqrt{2A}i$ & $\frac{\sqrt{2}}{\sqrt{A}}(2+3A)i$ & $\alpha$ &$-1-\frac{3A}{2}$\\
			I & $-3\sqrt{2A}i$ & $-\frac{\sqrt{2}}{\sqrt{A}}(2+3A)i$ & $\alpha$ &$-1-\frac{3A}{2}$\\
			J & $-\sqrt{3}\sqrt{-(2+3A)}$ & $\sqrt{3}\sqrt{-(2+3A)}$ & $\frac{6k\lambda-1}{3-6k\lambda}$ & - \\
			\noalign{\smallskip}\hline
		\end{tabular}
	\caption{Eigenvalues and quantities $q,\gamma$ for critical points of bouncing cosmologies}
\label{table2}}      
\end{center}
\end{table}
\begin{figure}[h!]
	\centering
	\subfigure[]{\includegraphics[width=0.325\textwidth]{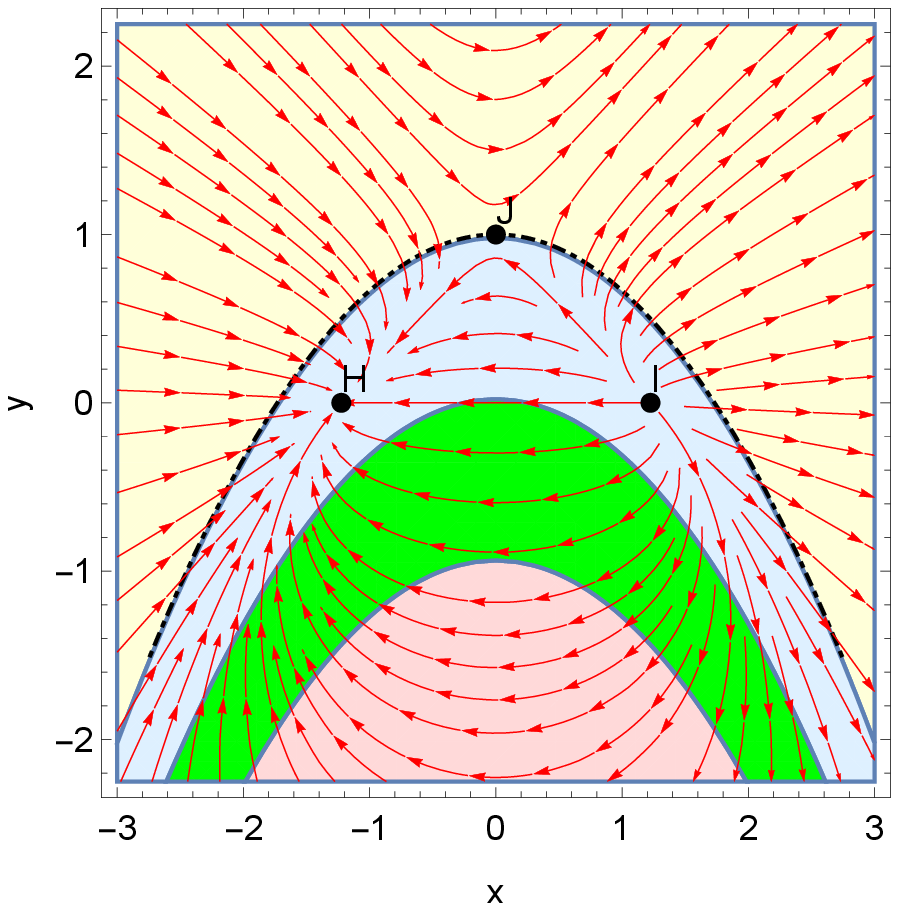}} 
	\subfigure[]{\includegraphics[width=0.325\textwidth]{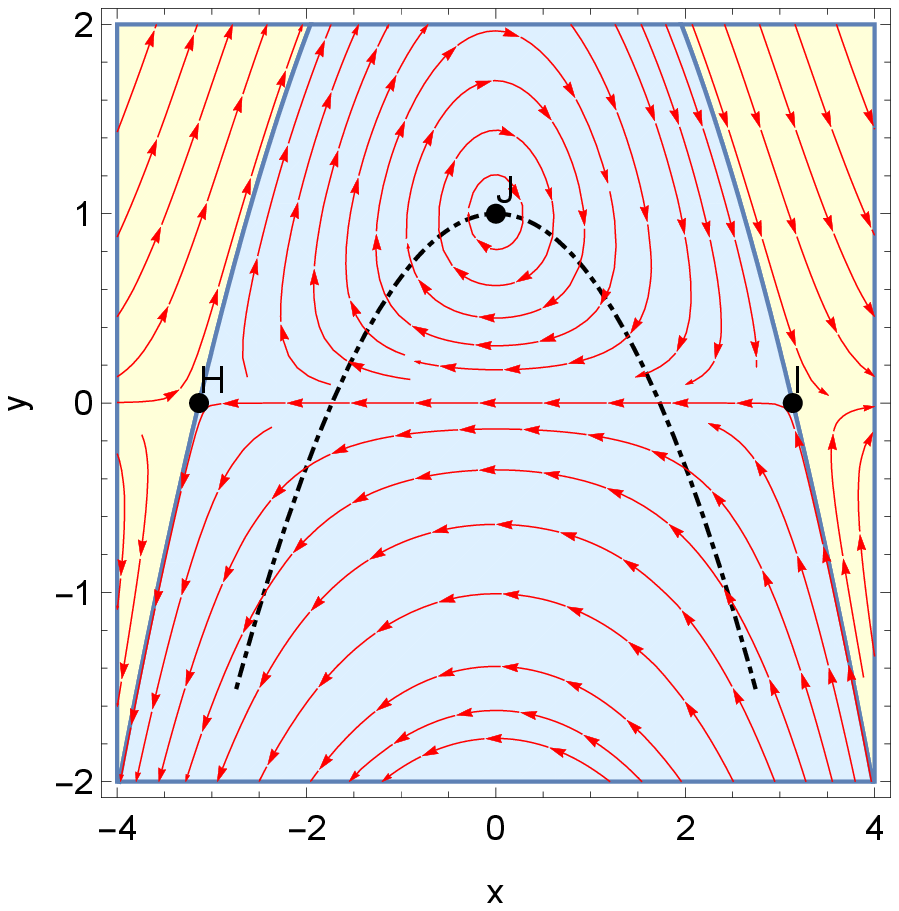}} 
	\subfigure[]{\includegraphics[width=0.325\textwidth]{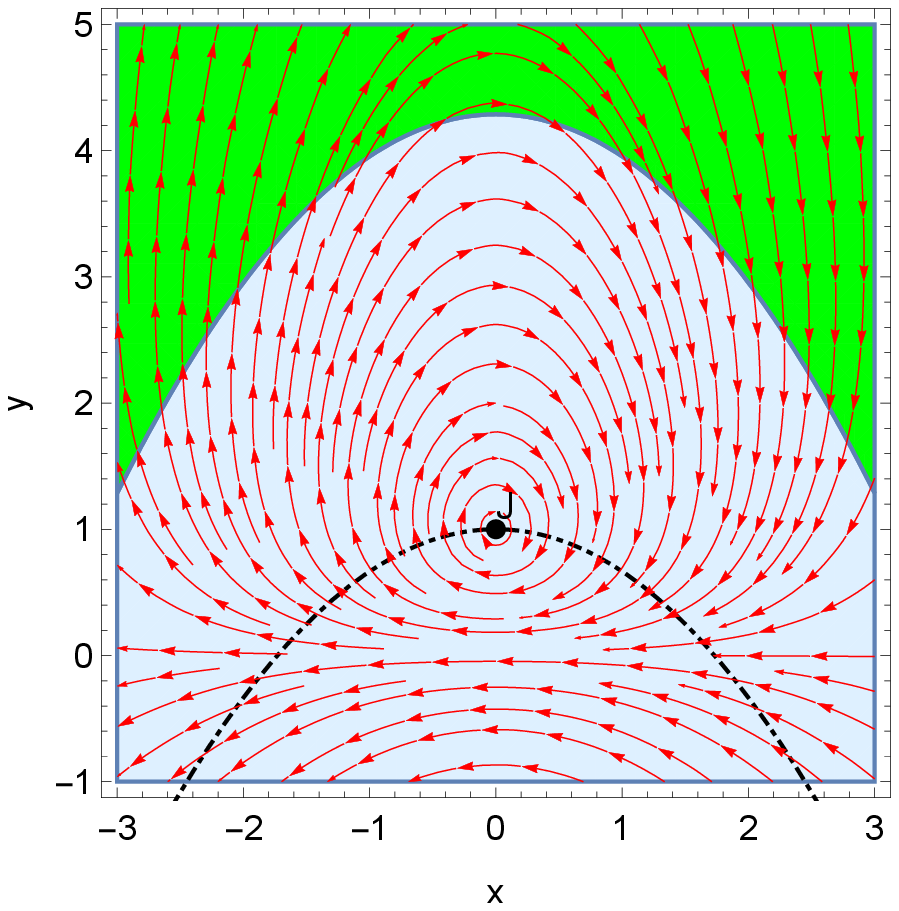}}
	\caption{Different views of phase space $(\bar{x},\bar{y})$ for (a) $A<-\frac{2}{3}$ (b) $-\frac{2}{3}<A<0$ (c) $A>0$ respectively}
	\label{fig4}
\end{figure}
For the autonomous system (\ref{eq21},\ref{eq22}), there will be three critical points given by $H(x=-\frac{1}{\Xi},y=0),I(x=\frac{1}{\Xi},y=0)$ 
and $J(x=0,y=1)$ where we denote $\Xi=\frac{\sqrt{(1+\alpha)(4k\lambda-1)}}{\sqrt{6k\lambda(1+\alpha)-2}}$. Various detail about these critical 
points and geometrical behavior in the $\bar{x}-\bar{y}$ plane has been summarized in Table (\ref{table2}) and Fig. (\ref{fig4}). In phase 
space diagrams $\gamma<-1,-1<\gamma<0,0<\gamma<1$ and $\gamma>1$ region have been highlighted by light-red, light-yellow, light-blue, green 
shades respectively. For $(\alpha,k,\lambda,A)$ equals to $\left( \frac{1}{3},1,0.163,-1.3238\right)$, $\left( 0,1,0.235,-0.20339\right)$ 
and $\left( 0.15,1,0.261,0.50828\right)$, we plot left, middle and right panel of Fig. (\ref{fig4}) respectively. At critical point $(x_0,y_0)$  
of the phase space, the scale factor takes the form $a=a_0 (t-t_0)^{-\frac{{\bar{x}_0}^2}{3(\bar{y}_0-1)}}$.\\
Point $H$ and $I$ will exist for $A<0$ only. Due to the existence condition, these points does not exhibit super-accelerated expansion, 
since we will need $A>0$. With $A<0$, we may get $q>-1$. In particular, $q<0$ for $A\in\left(-\frac{2}{3},0 \right)$, $q=0$ at $A=-\frac{2}{3}$ 
and $q>0$ for $A<-\frac{2}{3}$. Thus, we may conclude that these points will undergo accelerated
expansion with repeller/attractor nature and decelerated expansion with saddle nature. From left panel of Fig. (\ref{fig4}), we may observe 
that point $H$ and $I$ are acting as attractor and repeller respectively and in middle panel, these both points are saddle
in nature exhibiting decelerating universe. For $0\leq \alpha\leq 1$, we may get solutions
which may exhibit matter and radiation dominated universe at these point. The black curve in Fig. (\ref{fig4}) corresponds to $q=0$ curve 
except at point $J$ where $q$ is not defined due to zero Hubble parameter value.\\
Point $J$ will exist for all values of $\alpha$ and $k\lambda$. The point will be saddle point for $A<-\frac{2}{3}$ and center for 
$A>-\frac{2}{3}$. The trajectories in phase space revolve around the
center and does not converge to it, therefore the system will realize the oscillating
solutions of model.
\subsection{Analysis of critical points at infinity}
\begin{figure}[h!]
	\centering
	\subfigure[]{\includegraphics[width=0.325\textwidth]{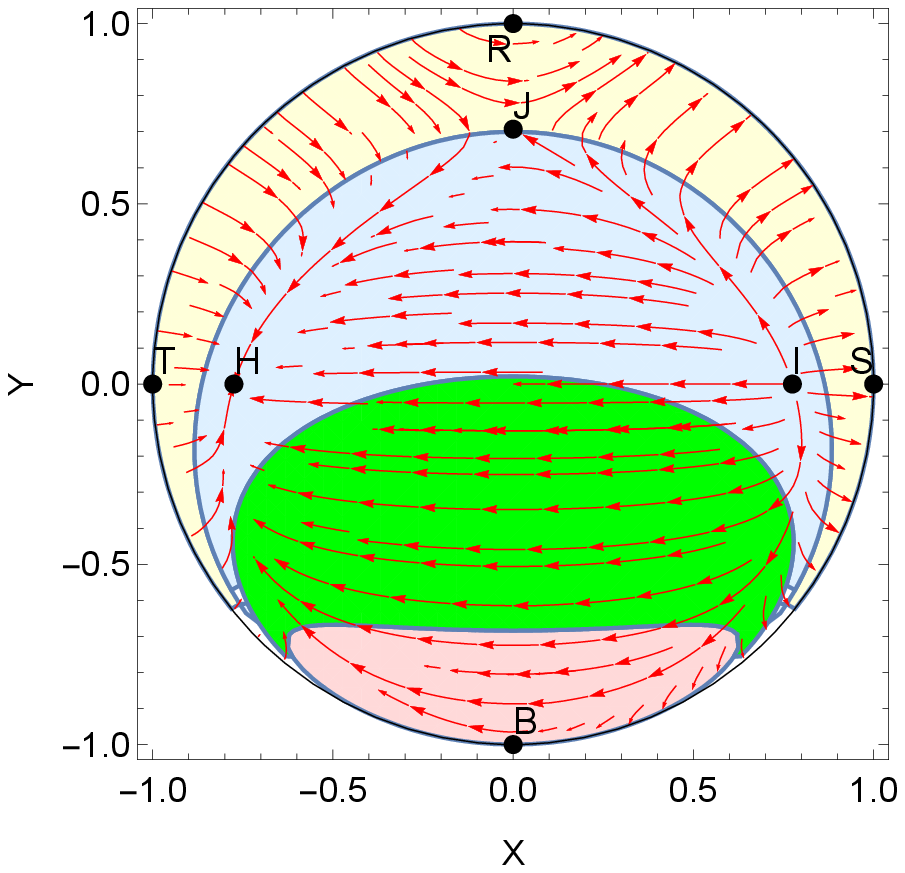}} 
	\subfigure[]{\includegraphics[width=0.325\textwidth]{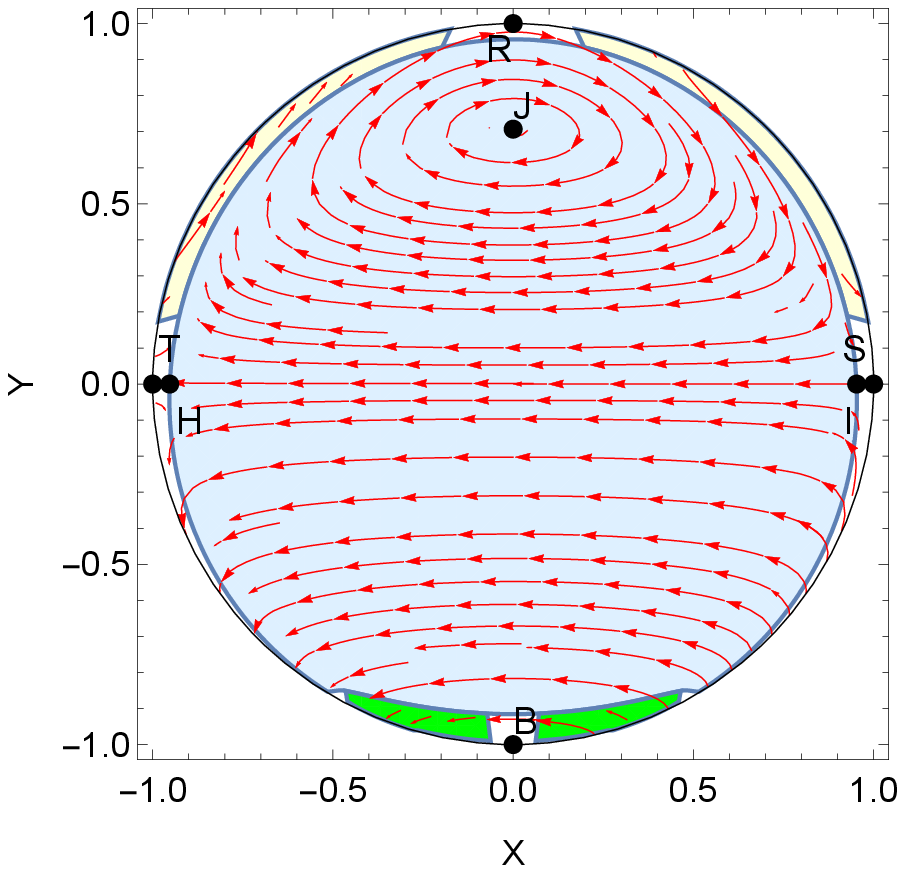}} 
	\subfigure[]{\includegraphics[width=0.325\textwidth]{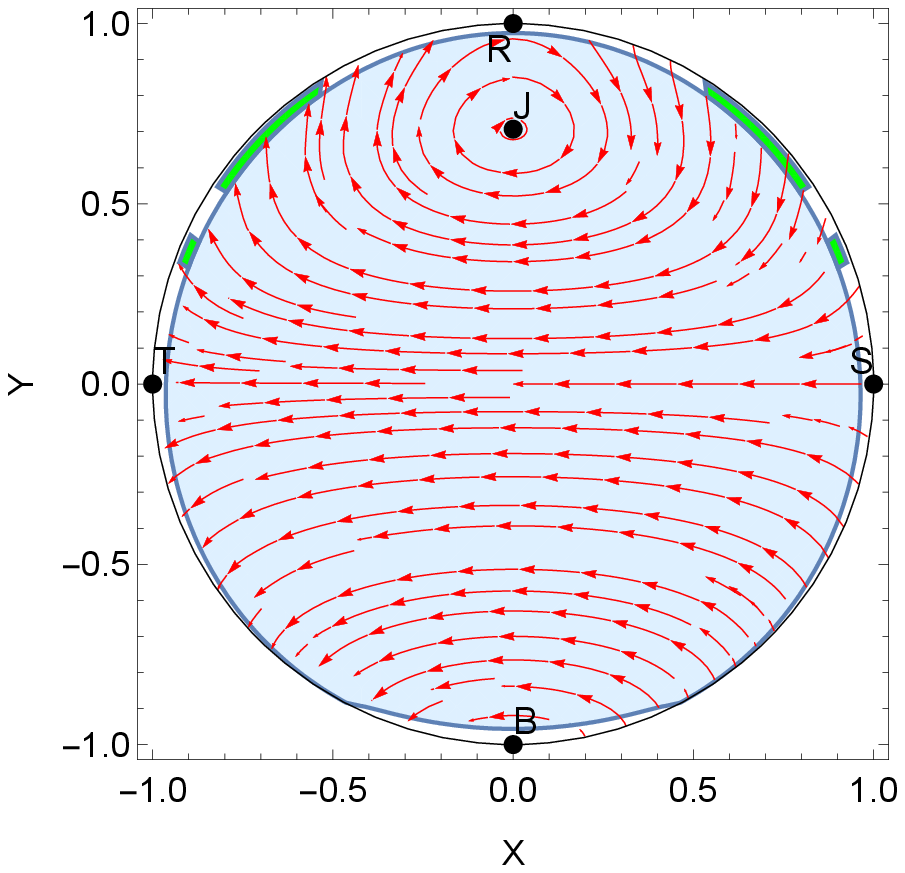}}
	\caption{Different views of phase space $(\bar{X},\bar{Y})$ for (a) $A<-\frac{2}{3}$ (b) $-\frac{2}{3}<A<0$ (c) $A>0$ respectively}
	\label{fig5}
\end{figure}

To understand the global dynamics of model, we may utilize the Poincare compactification
approach. The phase space in this model is having variables $\bar{x},\bar{y}$ and we
define $P=-\frac{3A}{2}{\bar{x}}^2+3\bar{y}-3$ and $Q=-(2+3A)\bar{x}\bar{y}$. By following the method stated
in the Appendix (\ref{append}), we write $P^{*}(\bar{X},\bar{Y},\bar{Z})=-\frac{3A}{2}{\bar{X}}^2+3\bar{Y}\bar{Z}-3{\bar{Z}}^2$ 
and $Q=-(2+3A)\bar{X}\bar{Y}$. Under the limit $\bar{Z}\rightarrow 0$, $\bar{X}Q_m-\bar{Y}P_m=-\left( 2+\frac{3A}{2}\right)\bar{X}^2\bar{Y}$ 
along with $\bar{X}^2+\bar{Y}^2=1$.  In this case $m=2$. So, the equilibrium points are $(\pm 1,0)$ and $(0,\pm 1)$. 
The phase space in variables $(\bar{X},\bar{Y})$ have been
given in Fig. (\ref{fig5}). The phase space exhibits the stability nature of critical points of
finite space along with the critical points near infinity. \\
We may use a fan-out map for study of the flow near equilibrium, along the circle of infinity for $\bar{X}>0$ and $\bar{Y}>0$ respectively \cite{47}. \\
Consider the points $(\pm 1,0)$, at these points $q=-1$ and $\gamma=-1$. In order to study an equilibrium at $\bar{Z}=0$ satisfying $\bar{X}>0$, 
we may map a point $(\bar{X},\bar{Y},\bar{Z})$ is mapped to $U=\frac{\bar{Y}}{\bar{X}},V=\frac{\bar{Z}}{\bar{X}}$. The differential 
equations governing the motion are
\begin{equation}
	U'=-\left(2+\frac{3A}{2}\right)U,\quad V'=\frac{3A}{2}V
	\label{eq24} 
\end{equation}
The equilibrium is at $(U,V)=(0,0)$ and the eigenvalues are $\lambda_1=2+\frac{3A}{2}$ and $\lambda_2=-\frac{3A}{2}$. The point have saddle 
nature for $A>0$ and $A<-\frac{4}{3}$. The point acts as
repeller for $-\frac{4}{3}<A<0$. Similarly, an equilibrium at $\bar{Z}=0$ satisfying $\bar{X}<0$ will also be a saddle point.\\
Consider the points $(0,\pm 1)$, at these points $q$ and $\gamma$ becomes indeterminate. Consider an equilibrium at $\bar{Z}=0$ satisfying 
$\bar{Y}>0$. A point $(\bar{X},\bar{Y},\bar{Z})$ is mapped to $U=\frac{\bar{X}}{\bar{Y}},V=\frac{\bar{Z}}{\bar{Y}}$. The differential 
equations governing the motion are
\begin{equation}
	U'=\left(2+\frac{3A}{2}\right)U^2,\quad V'=(2+3A)UV
	\label{eq25} 
\end{equation}
The equilibrium is at $(U,V)=(0,\delta)$, $\delta\in \mathbf{R}$ and the eigenvalues are $\lambda_1=0$ and $\lambda_2=0$. 
The equilibrium at $\bar{Z}=0$ satisfying $\bar{Y}<0$ will also have a similar nature. From Fig. (\ref{fig5}), we may observe that 
for selected value of model parameters, the points are saddle in nature.
\subsection{Energy conditions and the oscillating universe}
In the phase space diagrams of Fig. (\ref{fig4}) and (\ref{fig5}), curves have closed trajectories
around the center. It is the interesting feature of centers that the nearby trajectories
rotates about it, instead of converging to it and therefore realize oscillating solution(s) for the cosmological model. The oscillating 
solutions of the cosmological model may be termed oscillating universes. The feature of oscillating universe is that it avoids the 
singularity by replacing with a cyclical evolution from a previous existence. Therefore, we may utilize the centers to exhibit oscillating 
solutions in the phase space in cosmological modeling. The real part of eigenvalues of the linearized matrix will be zero for cyclic 
(non-singular) solutions of the model, as centers are marginally stable. An oscillatory universe experience a sequence of bounce and turnaround 
where it achieves minima and maxima in the scale factor \cite{36}. At the extremum in terms of scale factor of universe $\dot{a}_{t_i}=0$,
where $i=b,ta$ is for
bounce and turnaround, respectively. In FRW space-time, the conditions for bounce and turnaround may be written as \cite{36}
\begin{enumerate}
	\item $\dot{a}_{t_b}=0$ and $\ddot{a}>0$ whenever $t\in (t_b-\epsilon,t_b)\cup (t_b,t_b+\epsilon)$ for $\epsilon>0$.
	\item $\dot{a}_{t_{ta}}=0$ and $\ddot{a}<0$ whenever $t\in (t_{ta}-\epsilon,t_{ta})\cup (t_{ta},t_{ta}+\epsilon)$ for $\epsilon>0$.
\end{enumerate}
Equivalently, in terms of Hubble parameter, at extremum $H=0$ and in small region around the extremum, $\dot{H}>0$ and $\dot{H}<0$ for 
bounce and turnaround respectively. The bounce conditions in terms of dynamical variable $\bar{x}$ may be written by
\begin{enumerate}
	\item $\bar{x}>0$ corresponds to the expanding universe regime.
	\item $\bar{x}=0$ corresponds either to bounce, a re-collapse or a static universe.
	\item $\bar{x}<0$ corresponds to the contracting universe regime.
\end{enumerate}
For $\bar{x}=0$ case, we may use derivative to identify different scenarios \cite{27}
\begin{enumerate}
	\item $\frac{d\bar{x}}{d\bar{N}}>0$ corresponds to bounce.
	\item $\frac{d\bar{x}}{d\bar{N}}<0$ corresponds to a re-collapse.
	\item $\frac{d\bar{x}}{d\bar{N}}=0$ does not give enough information and one may use the high derivatives
	or analyze the neighboring phase space.
\end{enumerate}
From Eq. (\ref{eq3}) and (\ref{eq4}), the point-wise energy conditions at $t=t_i$ may takes the form
\begin{eqnarray}
	k\rho=-6k\lambda \dot{H}+3(1-2k\lambda)\frac{\kappa}{a^2}
	\label{eq26}\\
	k(\rho-p)=2(1-6k\lambda) \dot{H}+4(1-3k\lambda)\frac{\kappa}{a^2}
	\label{eq27}\\
	k(\rho+p)=-2 \dot{H}+2\frac{\kappa}{a^2}
	\label{eq28}\\
	k(\rho+3p)=6(2k\lambda-1) \dot{H}+12k\lambda \frac{\kappa}{a^2}
	\label{eq29}
\end{eqnarray}
In flat and open FRW model, NEC is satisfied for $k<0 \ (k>0)$ at and near the bounce (turnaround) point.\\
For closed FRW model, NEC is satisfied at bounce with $\dot{H}<\frac{1}{{a_b}^2}$ for $k>0$ and
NEC is always satisfied for $k>0$ at turnaround point.\\
From Eq. (\ref{eq5}), at the bounce or turnaround, we may write\cite{36}
\begin{equation*}
	\ddot{\rho}=3\beta(\rho+p)\dot{H}
\end{equation*}
where $\beta\equiv\frac{A}{1+\alpha}$. For $\beta>0$, we either have $A>0,\alpha>-1$ or $A<0,\alpha<-1$ and, for $\beta<0$, we either 
have $A>0,\alpha<-1$ or $A<0,\alpha>-1$. At the bounce
\begin{enumerate}
	\item Energy density will achieve it's minimum $(\rho_{min})$ whenever NEC is satisfied
	with $\beta<0$ or NEC is violated with $\beta>0$.
	\item Energy density will achieve it's maximum $(\rho_{max})$ whenever NEC is satisfied
	with $\beta>0$ and NEC is violated with $\beta<0$.
\end{enumerate}
At the turnaround:
\begin{enumerate}
	\item Energy density will achieve it's minimum $(\rho_{min})$ whenever NEC is satisfied
	with $\beta>0$ or NEC is violated with $\beta<0$.
	\item Energy density will achieve it's maximum $(\rho_{max})$ whenever NEC is satisfied
	with $\beta<0$ and NEC is violated with $\beta>0$.
\end{enumerate}
Satisfaction and/or violation of energy conditions near the extremum depend on the values of model parameters $\alpha,k\lambda$. The evolution 
of the non-singular universe may be identified based on the presence of a center in the system and the behavior of energy conditions. 
The null energy condition behavior will also affect the energy density value at the extremum of the oscillating universe. Point $J$ in Fig. (\ref{fig4}) 
and Fig. (\ref{fig5}) exhibit the oscillating universe for selected values of model parameters. Point $J$ exists in region $\bar{y}>0$ 
which means that in the FRW model with positive spatial curvature, we may have oscillating solutions exhibiting oscillating/cyclic universes.\\
At the extremum of non-singular (cyclic) universe, the conservation Eq. (\ref{eq5}) takes the form\cite{23} 
\begin{equation}
	p=\frac{1-3k\lambda}{3k\lambda}\rho+\frac{c}{3k\lambda}\label{eq30}
\end{equation}
The oscillating universe in the present Rastall gravity model will satisfy an affine equation of state in the small neighborhood 
of the extremum (bounce or turnaround point). Note that in Eq. (\ref{eq30}), we only use the definition of bounce or turnaround. This is an 
interesting aspect of Rastall gravity caused by the non-minimal
coupled nature of gravity. This aspect is opposite to the results of bouncing General Relativity models in addition to other minimally 
coupled bouncing models.
\section{Conclusions}
We investigate the Rastall gravity model with FRW line element by using dynamical system analysis for expanding and bouncing cosmologies. 
For the present model, we have found the critical points that may describe accelerating and decelerating eras representing dark energy and 
radiation/matter-dominated epochs.\\
In expanding models with $H>0$, we investigate for attractor critical point representing dark energy dominated solution. The component with 
negative pressure
(also called dark energy) becomes dominant only during late times to cause the current accelerated expansion of the universe. In the terminology of 
dynamical system and
its critical point, the stable critical point with accelerated expansion describes a dark energy solution. The model exhibits three eras of 
the universe, starting from a radiation-dominated stage, then a matter-dominated stage, and de Sitter behavior during late times. We have 
shown that the FRW universe is spatially flat for selected values of model parameters, having transitioned from a radiation-dominated phase 
at early times, followed by a matter-dominated era describing 
decelerating universe to an accelerating universe at late times. Point $O$ represents attractor behavior (with EoS parameter value $-1$) 
at late times where surrounding trajectories are attracted towards it, irrespective of initial conditions. The model can recover any cosmological era for different values of model parameters $(\alpha,k\lambda)$. 
This is a striking feature of the present model. The model also possesses a Milne solution in the FRW model with negative spatial curvature. We also 
investigate critical points near infinity for the model. Some of the points at infinity represent the viable cosmological features with 
saddle nature. In the Rastall gravity model with flat spatial section of FRW geometry \cite{51b}, the Rastall parameter $\lambda$ (with $k=1$) has been constrained to be in range 
	$-0.0007<\lambda<0.0001$ at $68\%$ CL from CMB+BAO data. The present qualitative analysis suggests to keep $\lambda>0$ with $k>0$ for keeping total matter density positive (and thus validity of WEC) during universe evolution in expanding cosmologies.\\
The Rastall gravity model may also describe cyclic models which go through alternating accelerating and decelerating phases. For different 
values of model parameters, we found the eigenvalues of critical points which represent centers exhibiting cyclic / oscillating solutions. 
In the present model, we consider barotropic fluid satisfying affine EoS in Rastall gravity framework to highlight different eras of universe. 
Model independent bounce and turnaround conditions when used in Rastall gravity also yields affine EoS at and near the bounce / turnaround, 
which is also an interesting aspect of the model. One may consider other forms of barotropic fluids along with matter and/or radiation 
component to examine the qualitative features of Rastall gravity.
\section*{Acknowledgments}
We are grateful to the anonymous reviewers for their perceptive and illuminating remarks, which have aided in the revision of work 
to enhance the quality of paper. A. Pradhan also thanks the IUCAA for providing facility and support under visiting associateship program.

\section*{Appendix: Poincare compactification}
\label{append}
We take the system
\begin{equation}
	x'=P(x, y),\quad y'=Q(x, y)
	\label{eq1a}
\end{equation}
Given a point on plane $(x,y)$, we may map this point uniquely to a point on the half sphere $S^{2+}=\{(X,Y,Z):X^2+Y^2+Z^2=1,Z\geq 0\}$. 
That is, each point $(x,y)\in \mathbf{R^2}$ will be projected, through the center of sphere, to a unique point in $S^{2+}$ in such a way that 
`infinity' in $(x,y)$ is now mapped at $Z=0$, that is, at the circle of infinity $X^2+Y^2=1$.\\
Under the map $x=\frac{X}{Z},y=\frac{Y}{Z}$, the above system (\ref{eq1a}) becomes
\begin{equation}
	X'=Z[(1-X^2)P-XYQ], \quad 	Y'=Z[-XYP+(1-Y^2)Q], \quad Z'=-Z^2[XP+YQ]
	\label{eq1b}
\end{equation}
where $P\left( \frac{X}{Z},\frac{Y}{Z}\right)$ and $Q\left( \frac{X}{Z},\frac{Y}{Z}\right)$ are now evaluated in terms of $\frac{X}{Z}$ 
and $\frac{Y}{Z}$. Regularize
the above function by defining $P^{*}(X,Y,Z)=Z^mP\left( \frac{X}{Z},\frac{Y}{Z}\right)$, $Q^{*}(X,Y,Z)=Z^mQ\left( \frac{X}{Z},\frac{Y}{Z}\right)$ 
where $m$ is the maximum degree of $P,Q$ and re-scaling $d\tau=Z^{1-m}dt$ the above system (\ref{eq1b}) becomes complete and valid on $S^{2+}$.\\
The complete system that allow us to study the (generally non-trivial) dynamics at
infinity is given by
\begin{equation}
	X'=-Y[XQ_m-YP_m], \quad Y'=X[XQ_m-YP_m]
	\label{eq1c}
\end{equation}
along with $X^2+Y^2=1$, where $P^{*}\rightarrow P_m$ and $Q^{*}\rightarrow Q_m$ as $Z\rightarrow 0$. There will be equilibrium points at 
infinity only for $XQ_m-YP_m=0$ along with $X^2+Y^2=1$ \cite{47,54,55}. Due to the $X^2+Y^2=1$ condition, these equilibria are of form $(X,Y,0)$. 
These equilibria will come in as antipodal pairs distributed along the circle of infinity. So,
it will be only necessary to investigate the flow along equilibria having $X>0$ and $Y>0$. In particular, they cannot both be zero \cite{54}. \\ 
The flow will be topologically equivalent (with direction reversed) at the antipodal points if $m$ is odd (even) as in the homogeneous 
case \cite{47}. Recall that $m$ is the
maximum degree of $P,Q$.
\subsection*{Flow near equilibria}
The flow near equilibria, along the circle of infinity may be studied by using a fanout map\cite{47} for $X>0$ and $Y>0$ respectively.\\
Consider an equilibrium at $Z=0$ satisfying $X>0$. Point $(X,Y,Z)$ is mapped to $U=\frac{Y}{X},V=\frac{Z}{X}$. Differential equation governing 
the motion are\cite{47}
\begin{equation}
	U'=V^m\left(Q_m\left(\frac{1}{V},\frac{U}{V}\right) -UP_m\left(\frac{1}{V},\frac{U}{V}\right) \right)  
	\label{eq1d}
\end{equation}
\begin{equation}
	V'=-V^{m+1}P_m\left(\frac{1}{V},\frac{U}{V}\right) \label{eq1e}
\end{equation}
In order to study an equilibrium, at $Z=0$ satisfying $Y>0$, we may map a point $(X,Y,Z)$ to $U=\frac{X}{Y},V=\frac{Z}{Y}$. 
The differential equation governing the motion are \cite{47}
\begin{equation}
	U'=V^m\left(P_m\left(\frac{U}{V},\frac{1}{V}\right) -UQ_m\left(\frac{U}{V},\frac{1}{V}\right) \right)  
	\label{eq1g}
\end{equation}
\begin{equation}
	V'=-V^{m+1}Q_m\left(\frac{U}{V},\frac{1}{V}\right) \label{eq1h}
\end{equation}
For complete elaboration of steps involved in above equations, one may refer to Perko\cite{47}.


\end{document}